# Stability of Two-Dimensional Iron-Carbides Suspended across Graphene Pores: First-principles Particle Swarm Optimization

Yangfan Shao, Rui Pang, and Xingqiang Shi[*]

*Department of Physics, South University of Science and Technology of China, Shenzhen 518055, China*

*E-mail: shixq@sustc.edu.cn

**ABSTRACT:** Inspired by recent experimental realizations of two-dimensional (2D) metals and alloys, we theoretically investigate the stability and electronic properties of monolayer (ML) Fe-C compounds and pure Fe. According to our and others theoretical results, the experiment [Science 343, 1228 (2014)] proposed ML pure Fe *square*-lattices embedded in graphene (Gr) pores are energetically unstable compared to that of the Fe *triangular*-lattices in Gr. To solve the above contradiction, we search for the stable structures of ML Fe-C with various Fe to C ratios (as a generalization of ML Fe in Gr) using *ab initio* particle swarm optimization technique. A $Fe_1C_1$ *square*-lattice embedded in Gr is found. We propose and demonstrate that the square-lattices observed in the experiment were iron-carbides (Fe-C) but not pure Fe from the square-lattice shape, Fe-Fe lattice constant and energetic considerations. Note that the coexistence of C with Fe cannot be excluded from the experiment. More importantly, we find a lowest energy and dynamically stable structure, ML $Fe_2C_2$ with Fe atoms form distorted square lattices. High spin polarization around the Fermi level is predicted for different 2D Fe-C structures due to significant orbital hybridization between C and Fe.

# INTRODUCTION

Great enthusiasm for finding new two-dimensional (2D) and quasi-2D materials has been stimulated since the discovery of graphene (Gr). Low-dimensional materials exhibit lots of excellent properties [1-2], and low-dimensional *magnetic* materials are keys to develop the next generation storage devices with higher storage density and further miniaturization[3]. The structure of materials is one of the most important determinants of their properties. Determination of materials' structures is a high priority for studying their properties. Recently, the evolutionary algorithms have been used for theoretical prediction of crystal structures, many stable crystal structures were successfully predicted from evolutionary algorithms in combination with *ab initio* calculations [4-6].

For the study of 2D single-atom-thick metallic materials and their alloys, Duan *et al.* reported the fabrication of poly(vinylpyrrolidone)-supported single-layer rhodium (Rh) nanosheets using a facile solvothermal method[7]. Koskinen *et al.* predicted the existence of an atomically thin, free-standing 2D liquid phase of gold (Au), which is enabled by its exceptional planar stability due to relativistic effects[8]. Saleem *et al.* prepared a free-standing Pt-Cu alloy material down to one-atom-thick [9].

For the study of 2D single-atom-thick iron (Fe) based structures, Wang *et al*. demonstrated that graphene edges could prevent the collapse of Fe from a 2D single-atomic monolayer (ML) to a 3D cluster by density functional theory calculations[10]. They found that the lower dimensionality of ML Fe remarkably altered its orbital overlap, and hence the ML Fe magnetic moment is increased to 2.68 $\mu_B$ per atom, as compared to bulk Fe of 2.20 $\mu_B$. The ML Fe in contact with Gr is in a 2D *close-packed triangular-lattice*. Zhao *et al.* observed a 2D *less-close-packed square-lattice* structure suspended across perforations in graphene, they proposed that the embedded square-lattice to be made up of pure Fe[11]. They observed experimentally that the lattice constant of the square structure was 2.65±0.05 Å, while their first-principles calculation results suggested that the lattice constant for ML Fe was only about 2.35 Å. Thomsen *et al.* also studied the stability of monolayer Fe using *ab initio* calculations[12], and found that the ferromagnetic *triangular*-lattice of Fe membrane was the most stable. So, there's a gap in the interpretation of the experimentally observed less-close-packed *square*-lattice structures -- all theoretical results for pure ML Fe prefer close-packed *triangular*-lattices.

In the current work, we theoretically study the stability and electronic properties of single-atom-thick iron and iron carbides (Fe-C). Iron carbides are considered since the coexistence of C with Fe cannot be excluded from the Fe-in-Gr structures in experiment[11]. Firstly, the stability of Fe square-lattice in graphene pores are studied by *ab initio* total energy structural relaxations -- the Fe square lattices changed to triangular lattices after relaxation. Furthermore, we search for the stable structures of 2D Fe-C compounds with various Fe to C ratios (including Fe in Gr) with the help of particle swarm optimization (PSO) technique. A monolayer $Fe_1C_1$ structure with a square-lattice (Fe atoms in $Fe_1C_1$ also form a square lattice) suspended in graphene is found by our PSO global structural search. We then propose that the embedded structure in Gr to be iron carbides (Fe-C) instead of pure iron from the square lattice-shape, Fe-Fe lattice-constant and energetic considerations. More importantly, we find a lowest energy monolayer $Fe_2C_2$ structure (Fe atoms themselves form distorted square lattices), which is dynamically stable. Finally, the spin-polarized electronic structures of the $Fe_2C_2$ membrane are calculated both with and without hybrid density functionals.

This paper is organized as the following. We begin with computational details in the next Section. In the third Section, first the results of pure Fe monolayers in free-standing and graphene-embedded configurations are presented; and then the results of iron carbide monolayers in free-standing and graphene-embedded configurations. Finally a summary is given.

## ■ COMPUTATIONAL METHODS

Spin-polarized density functional theory (DFT) calculations were performed using the Vienna ab initio simulation package (VASP)[13-14]. Projector augmented wave (PAW) potentials were used to describe the core electrons[15]. The valance electron orbitals were expanded using plane waves with a kinetic energy cutoff of 500 eV. The Perdew-Burke-Ernzerhof generalized gradient approximation (PBE-GGA) exchange and correlation functional was used for structural relaxation and electronic calculations[16]. For electronic property calculations of $Fe_2C_2$, both PBE and Heyd-Scuseria-Ernzerhof screened hybrid functional (HSE06)[17] were used. For density of states (DOS) calculations, Gaussian smearing of 0.05 eV was used. The 2D cell Monkhorst Pack *K*-point sampling was 15×15. The vacuum layer was 10 Å in all calculations.

To find the global minimum structure, we adopted 2D particle swarm optimization (PSO) algorithm as implemented in the CALYPSO code[18-19], which is interfaced with VASP. We first generated a set of random 2D structures with randomly chosen symmetries among the 17 plane-space-groups. The minimal interatomic distance constraint of 1.3 Å was used to make the random structures reasonable. The random structures were optimized by VASP to find the local minimums of lattice energies. Then 60% of total structures for the next generation were produced based on PSO operation and the rest of new structures were generated randomly. The lowest energy structures were obtained by iteration the above procedures.

To examine the dynamic stability of our predicated new structures, we calculated their phonon dispersion curves using the Phonopy program[20], which is interfaced with VASP. We used 5×5 supercells in the adopted finite displacement method[21] to simulate phonon dispersion curves. The kinetic-energy cutoff was 750 eV for phonon calculations. In addition, magnetic properties and electron localization function (ELF) analyses [22-23] were used to gain insight to Fe-C interactions and bonding properties for the new predicted iron carbides.

## ■ RESULTS AND DISCUSSIONS

**A. IRON MONOLAYERS**

**Monolayer Fe.** We first study the stability and magnetic properties of free-standing Fe monolayers to get a good research foundation for the Fe-in-graphene structures. Both square and triangular Fe monolayers in ferromagnetic (FM) and antiferromagnetic (AFM) ordering are considered. Firstly, we analyze monolayer Fe square- and triangular-lattices in FM ordering. For the FM square structure, the equilibrium lattice constant is 2.31 Å (see Fig. 1), while it is 2.41 Å for FM triangular lattice. Compared to the square-lattice, the triangle lattice is lower in energy by 0.32eV per Fe atom. The triangular lattice is more stable, which is in good agreement with Refs.[10, 12]. The spin magnetic moments are 2.66 $\mu_B$ and 2.63 $\mu_B$ in square and triangular lattices, respectively. Secondly, to consider the AFM ordering, we adopted a ($\sqrt{2}\times\sqrt{2}$)R45º supercell to do the AFM simulation of the square lattice. The AFM square-lattice is less stable than the FM one by 0.25 eV (Figure 1). The triangular "AFM" lattice is simulated with a ($\sqrt{3}\times1$) rectangular supercell. The "AFM" ordering is less stable by 0.15 eV. So, the FM triangular lattice is the most stable compared to "AFM" triangular, FM square, and AFM square, in the order from most stable to most unstable; this order is in good agreement with the calculated results in Ref. [12].

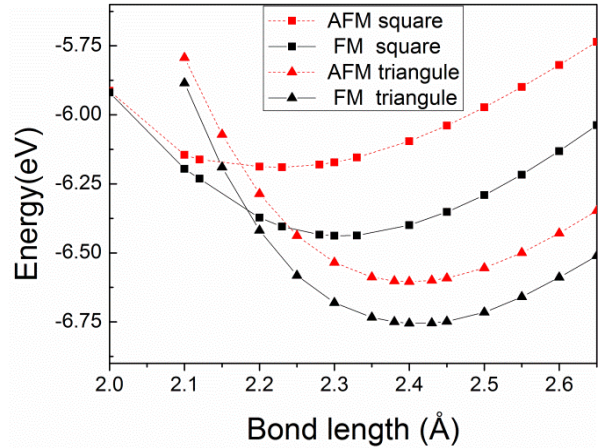

Figure 1. Potential energy surfaces of square- and triangular-lattice iron membranes in AFM and FM orderings.

**Monolayer Fe across Graphene pores.** Although for free-standing Fe membranes the most stable structure is triangle, one may suspect that the Fe *square*-lattice may be more stable than the triangular one when it embedded in graphene pores, as was proposed in the transmission electron microscopy (TEM) study[11]. We constructed *square* monolayer Fe embedded in graphene pores according to those proposed in the experiment[11]. Based on the scanned lattice constant for FM square Fe in the above Subsection, the lattice constant is adopted to be 2.31 Å in constructing initial structures for structural optimizations. There are two kinds of edges for graphene -- the zigzag and armchair edges. There are also two kinds of edges for monolayer square Fe along the (100) and the (110) directions. Three types of Fe monolayers in Gr pores can be constructed based on the two Fe edges: namely both Fe (110) and Fe (100) as edges (structure S1), only Fe(110) as edges (S2), and only Fe(100) as edges (S3). The initio Fe edges in structure S1 are Fe (110) alternating with Fe (100), forming an octagon as sketched in Figure 2a. However, the two Fe edges cannot align the two Gr edges simultaneously, due to the 15° small orientation mismatch between the edges of Fe and Gr: because the intersection angle between the two Fe edges is 45°, while that between the two Gr edges is only 30°. For example, as shown in Fig. 2b, if the Fe (110) edge aligns the Gr zigzag edge in the horizontal direction, then there's a 15° misalignment between the Fe (100) and the Gr armchair edges. Due to the 15° small orientation mismatch, to make reasonable Gr-Fe contacts, one cannot use exactly the same number of atoms in constructing the above mentioned three types of structures, S1, S2 and S3. For each type of structures we constructed small size, medium size and large size structures. A total of nine structures are relaxed. These structures after relaxation are shown in Figures 2d-f.

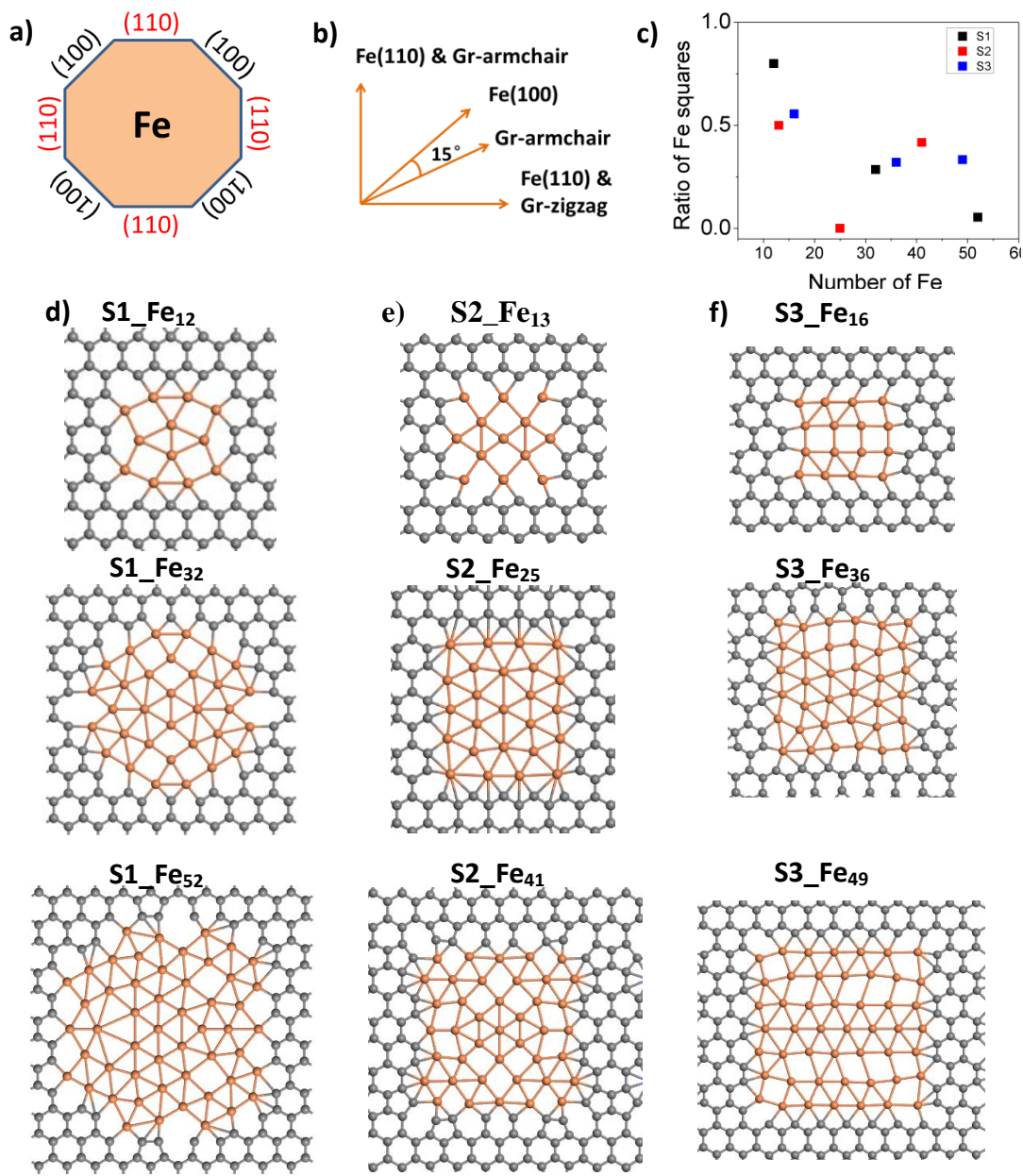

Figure 2. Optimized geometries of initial Fe squares embedded in graphene pores; most of the Fe squares changed to triangular after relaxation. (a) Sketch map of the two Fe edges in structure S1, forming an octagon; (b) misalignment between graphene and Fe edges (see text). (c) The ratio of survived Fe squares after relaxation. Structure types S1 (d), S2 (e), and S3 (f) along the Fe edges of both (110) and (100) directions, (110) only, and (100) only, respectively. In the labels Fe$_n$, n denotes the number of Fe atoms in graphene pores. Color codes: gray for C and orange for Fe.

All the relaxed structures are in 2D strictly, namely all atoms of Fe and C atoms are in the same plane, which is different to Ref.[12], where the Fe atoms are at different heights due to the fixed or nearly fixed number of Fe atoms. The Fe *square*-lattices are destroyed in graphene pores -- after relaxation a majority of the squares changes to triangles. To quantify how many squares changed to triangle, we classify the lattice to square when the Fe bond angle is in the range of 75° to 105°, otherwise it is classified to triangle. Most of the squares changed to triangles. The ratios of Fe *square*-lattice survived are less than 50% for medium- and large-size structures, as counted in Figure 2c. The octagonal edges of S1 (Figures 2a and 2d) are in good agreement with the experimental proposed graphene pore shapes[11], so we focus on type S1 structures for the ratio of survived squares. And, the experimental observed structures[11] corresponding to our simulated medium to large structures, namely S1_$Fe_{32}$ and S1_$Fe_{52}$. From these two relaxed S1 structures (Figure 2d, also refer to Figure 2c) we can see that only 28.6% and 5.0% squares survive for S1_$Fe_{32}$ and S1_$Fe_{52}$, respectively.

From the above results that most squares changed to triangles after relaxation, one can conclude that the suspicion that the Fe squares in graphene pores maybe more stable than triangles is incorrect. In the Gr-embedded structures, the Fe-Fe bond lengths are influenced compared with the free-standing structures due to the interaction with Gr at edges (the Gr-Fe contact interfaces); the Fe-Fe bond lengths of the Gr-embedded structures are in a range of about 2.1 to 2.6 Å. The average bond length is about 2.3 Å, much smaller than that of 2.65 Å in the experiment[11].

The average spin magnetic moments of Fe membranes are listed in Table I. The spin magnetic moments for monolayer Fe embedded in graphene are enhanced as compared to that of bulk Fe. The results also show that the spin magnetic moments of Fe at edges that contacted with Gr are reduced relative to pure ML Fe, due to the orbital hybridization and spin-polarized charge transfer between carbon and iron[24].

TABLE I. Average spin magnetic moments (Mag.) in $\mu_B$ of ML Fe embedded in Gr. S1, S2, and S3 are the structure types as shown in Fig. 1. Numbers without (with) brackets are the spin magnetic moments of all Fe (Fe at edges that bonds to C).

| S1 | Mag. ($\mu_B$) | S2 | Mag. ($\mu_B$) | S3 | Mag. ($\mu_B$) |
|---|---|---|---|---|---|
| $Fe_{12}$ | 2.82 (2.49) | $Fe_{13}$ | 2.62 (2.29) | $Fe_{16}$ | 2.49 (2.21) |
| $Fe_{32}$ | 2.64 (2.31) | $Fe_{25}$ | 2.67 (2.45) | $Fe_{36}$ | 2.39 (2.10) |
| $Fe_{52}$ | 2.73 (1.83) | $Fe_{41}$ | 2.34 (1.66) | $Fe_{49}$ | 2.86 (2.44) |

**B. IRON-CARBIDE MONOLAYERS**

To solve the contradiction between TEM image proposed pure Fe square lattices in Gr[11] and our total energy relaxations which prefer triangular ones, and to take fairly complete Gr-Fe (and more general C-Fe) structural configurations into considerations, we adopt a general global optimization method to search for lowest-energy 2D structures by the particle swarm optimization algorithm. The stable and metastable structures are predicted by PSO at given chemical compositions and stoichiometry for 2D layered structures. We consider the 2D Fe-C compounds with various Fe concentrations and then the compounds are divided into C-rich and Fe-rich cases.

**C-rich cases.** For the C-rich compounds, the structures of $(FeC_m)_n$ are predicted by PSO, where m = 2, 3 and n = 1 to 4 (for a set of fixed m and n, there are n Fe atoms and m*n C atoms in a simulation cell]. For different $(FeC_m)_n$, some similar configurations are predicted; among these configurations two most stable structures are obtained -- triangular-Fe-in-Gr and $Fe_1C_1$-in-Gr as shown in Figure 3a and b. The two structures in Figures 3a and b are predicted from both $(FeC_2)n$ and $(FeC_3)n$ cases for n = 2, 3, 4. Triangular-Fe-in-Gr is the lowest energy structure and it is consistent with the above results of triangle iron ML appears in graphene. However, the relative energy difference is only 0.06eV per atom between triangular-Fe-in-Gr and $Fe_1C_1$-in-Gr. Although triangular-Fe-in-Gr is the lowest energy structure, the energy difference is small between these two structures so that $Fe_1C_1$-in-Gr can also exist in experiment[11]. The average Fe-Fe bond length is 2.57 Å for $Fe_1C_1$-in-Gr; in the horizontal direction the Fe-Fe distance is 2.51 Å, which is shorter than that in the vertical direction of 2.62 Å (Figure 3b), due to the constraints in the horizontal direction by the smaller Gr lattice. So the Fe-Fe bond length of free-standing $Fe_1C_1$ should be larger than that in the $Fe_1C_1$-in-Gr structure; and this is confirmed in Figure 5a in Fe-rich cases -- the Fe-Fe distance is 2.65 Å in free-standing $Fe_1C_1$.

It is very difficult to observe C atoms near Fe in TEM due to the large difference in contrast between C and Fe.[12] We suspect that the observed structure actually is iron carbide but not pure iron from lattice constant and lattice size considerations. Only the $Fe_1C_1$-in-Gr structure can explain the TEM images observed *square*-lattice Fe, for both the square lattice shape and Fe-Fe distance of about 2.6 Å. Hence, in the following we focus on $Fe_1C_1$-in-Gr-like structures, i.e. Fe of Fe-C in

square- or nearly square-lattices.

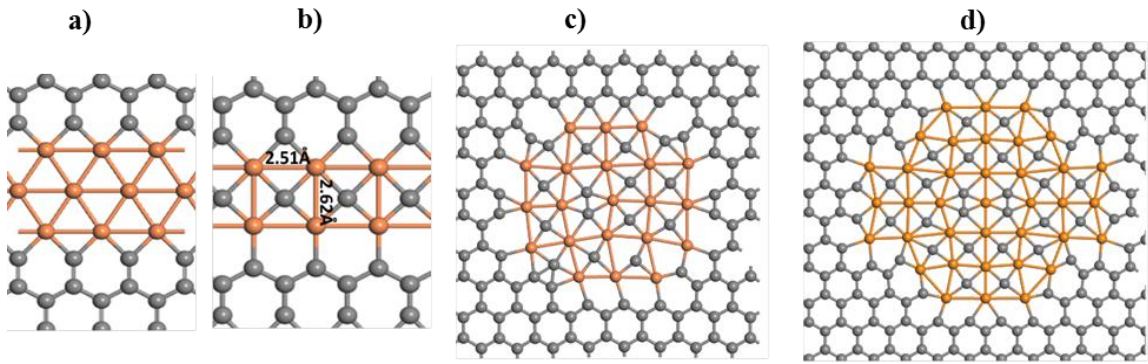

Figure 3. Low energy 2D structures of Fe-C compounds in C-rich cases: triangular-Fe-in-Gr (a) and $Fe_1C_1$-in-Gr (b) from 2D PSO algorithm. Base on $Fe_1C_1$-in-Gr, Fe-C compounds embedded in Gr "octagon" pores were constructed with two different pore sizes, the relaxed structures are shown in (c) for a medium size $Fe_{21}C_{16}$-in-Gr structure, and in (d) for a large size $Fe_{37}C_{32}$-in-Gr structure. Color codes: gray for C, orange for Fe.

Based on the $Fe_1C_1$-in-Gr structure in Figure 3b, we construct two monolayer Fe-C compounds embedded in Gr "octagon" pores with different sizes (Figure 3c and d). We relax the constructed structures and find that they remain in exactly 2D after relaxation (Figure 3c and d show the relaxed structures). Fe in the relaxed structures keep at their square-lattice positions, namely Fe-Fe itself in $Fe_1C_1$ form square lattices, and the lattice shapes are different from the pure-iron-triangular-lattices in graphene perforations (Figure 2). The irons in iron-carbides can keep in square-lattice sites due to the involving of carbons, while they changed to triangles for pure Fe in Figure 2. The large size $Fe_{37}C_{32}$-in-Gr structure (Figure 3d) structure has a $C_2$ symmetry. From the optimized structures, one sees that the square-lattice can survive in graphene pore for Fe-C. The bond lengths of Fe-Fe are in the range of about 2.45 to 2.85 Å, and the average Fe-Fe bond length is extremely close to the experimentally observed lattice constant of 2.65 Å, while for pure Fe it is only about 2.3 Å (Figure 2). Fe-Fe alone form square-lattices in the Fe-C structures. So, Figure 3c and d fit to TEM experiments[11] extremely good for both the nearly square-lattice shape and the average Fe-Fe distance of 2.65 Å.

There are Gr zigzag and armchair edges, and Fe (110) and (100) edges in structures of $Fe_{21}C_{16}$-in-Gr and $Fe_{37}C_{32}$-in-Gr (Figure 3c and d), so they can be considered as the typical structures in comparison to the experimental observed ones. We further study the electronic and magnetic properties for the typical structures. According to Bader charge analysis[25-26], Fe loses 0.79 electrons per atom. The charge transfer between Fe and C leads to the spin magnetic moment reduction of the iron[27]. The average spin magnetic moment of Fe is 1.60 μB in $Fe_{21}C_{16}$-in-Gr and 1.32 μB in $Fe_{37}C_{32}$-in-Gr, respectively. So the spin magnetic moments of Fe in ML Fe-C are smaller than that in ML pure Fe and bulk Fe.

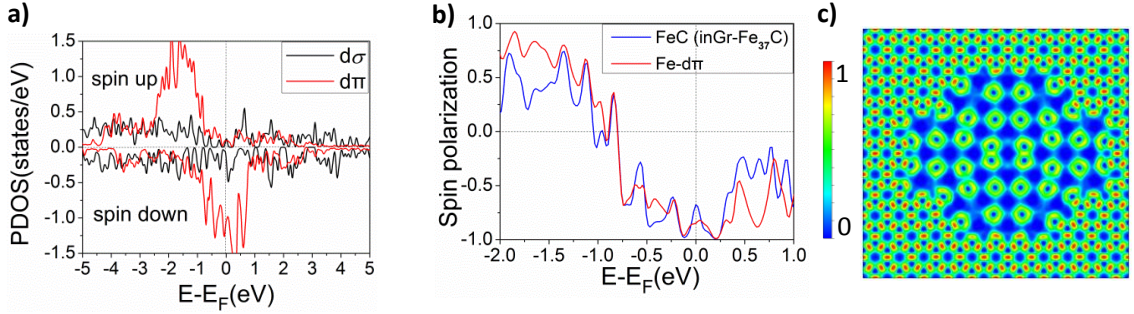

Figure 4. Electronic properties of ML Fe-C (Fe atoms and the C atoms enclosed by Fe) in $Fe_{37}C_{32}$-in-Gr (structure of Figure 3d). (a) PDOS of Fe-C with 0.05 eV Gaussian broadening, (b) spin polarization ratio of Fe-C and Fe out-of-plane $d_\pi$ orbitals, and (c) electron localization function (ELF) plot of $Fe_{37}C_{32}$-in-Gr.

In Figure 4a, the spin-polarized PDOS of the Fe-C (Fe atoms and the C atoms enclosed by Fe) in $Fe_{37}C_{32}$-in-Gr (structure of Figure 3d) are calculated to probe the effects of interaction between Fe and C. As shown in Figure 4a, the Fe-C in $Fe_{37}C_{32}$-in-Gr shows stronger spin-polarization. Compared to bulk Fe, the Fe-$d$ peaks are broadened in $Fe_{37}C_{32}$-in-Gr. Note that the Fe $d$ orbitals are decomposed to in-plane component $d_\sigma$ and out-of-plane component $d_\pi$ in Figure 4a. The corresponding Fe-C total spin polarization ratio is given in Fig. 4b, and the spin polarization ratio is defined as SPR($E$) = [DOS$_\uparrow$($E$)–DOS$_\downarrow$($E$)] / [DOS$_\uparrow$($E$)+DOS$_\downarrow$($E$)]. The spin polarization around Fermi level is increased as compared to bulk Fe. There is strong spin polarization for Fe-C in $Fe_{37}C_{32}$-in-Gr, up to 100% at ±0.2 eV around Fermi level. And the spin-polarization of the Fe conduction electrons $d_\pi$ is more significant.

The electron localization function (ELF) plot is used to assess the strength of the electron localization and it provides further information about the interaction and bonding. In order to give more comprehensive analysis of the Fe-C interaction, we simulated ELF plots of $Fe_{37}C_{32}$-in-Gr, as shown in Figure 4c. ELF is significantly distributed around the center of C-C bond of graphene, indicating the presence of the strong C-C covalent bonds. Moreover the delocalized Fe orbitals can be seen in the ELF plots, where the localization of electrons around the Fe is not strong (compared to C-C). There are less localized electrons around Fe atoms. It is interesting to note that the dim (blue) features in the ELF plot show the Fe-Fe square lattices.

**Fe-rich cases.** For the Fe rich case, the structures of $(FeC)_n$ (n = 1 to 5), $Fe_2C$, $Fe_3C$, $Fe_4C$ and $Fe_5C_2$ are searched via PSO. The structures of Fe-rich compounds transform to more than ML in thickness for Fe to C ratios larger than one. The searched structures can maintain 2D only when the ratios of Fe to C are not larger than one. From the experimentally obtained TEM micrographs and from comparing the intensity profiles between the simulated images and the experimental images, the ML thickness of Fe membranes can be determined in Ref. 11. Our current work concentrates on ML structures; hence we ignore the Fe-rich Fe-C compounds more than ML in thicknesses.

For $(FeC)_n$, for n = 1, a structure of $Fe_1C_1$ is found (Figure 5a); while for n = 2 to 5, another type of more stable structures, $Fe_2C_2$, are found (Figure 5b). Ryzhkov *et al.* studied the geometry, electronic structure, and magnetic ordering of iron-carbon clusters [27-28]. They found that the planar structure of $Fe_2C_2$ cluster is more stable than its 3D isomers. And they found the lowest energy $Fe_2C_2$ cluster is in a structure similar to our predicted $Fe_2C_2$ periodic structure.

The total energy of $Fe_2C_2$ is lower than $Fe_1C_1$ by 0.83 eV per $Fe_1C_1$, or lower by 0.37 eV than that of the separate systems of monolayer Fe and Gr.[12] So, $Fe_2C_2$ is the most stable structure than both $Fe_1C_1$ and the separate systems of ML Fe and Gr. The average Fe-Fe distance in $Fe_2C_2$ is 2.72 Å, which is close to the experimentally observed value. In Figure 3c and d, there is a trend of $Fe_1C_1$ changing to $Fe_2C_2$ at the center of Gr pores for both $Fe_{37}C_{32}$-in-Gr and $Fe_{21}C_{16}$-in-Gr. Fe itself in $Fe_2C_2$ is also in close-to-square arrangement. Based on the above analysis of lattice size, lattice shape, and total energies, we conclude that the square lattice in Gr observed in TEM images may be a mixture of (or something between) $Fe_2C_2$ and $Fe_1C_1$. Note that C atoms near Fe cannot be excluded in TEM images due to the large difference in contrast. Finally, it is very interesting to note that the $Fe_2C_2$ is extended to a "quasicrystal"-like structure (Figure 5b).

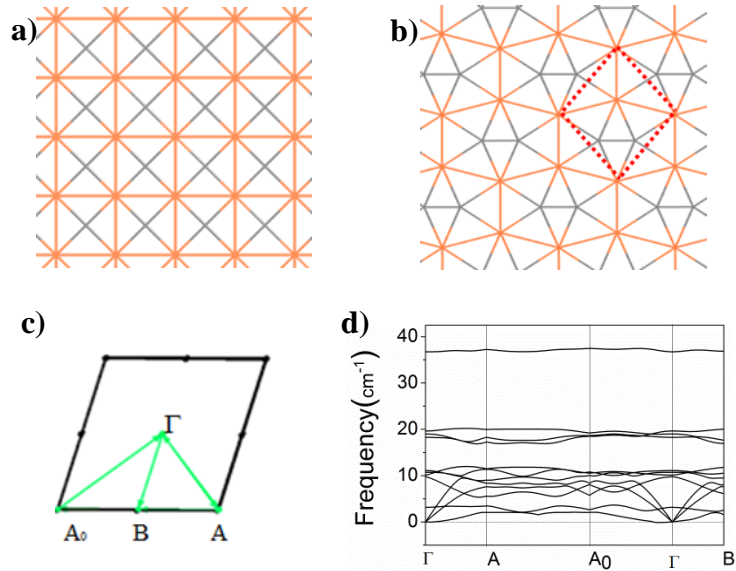

Figure 5. The $Fe_1C_1$ (a) and the $Fe_2C_2$ (b) compounds; $Fe_2C_2$ is the lowest energy 2D structure predicated from 2D PSO algorithm; the structures are shown with lines; gray for C and orange for Fe. (c) The first Brillouin zone and *k*-path of $Fe_2C_2$, and (d) the phonon dispersion curves of $Fe_2C_2$.

To check the dynamic stability of ML iron carbides and irons, we simulate their phonon dispersion curves. We checked the dynamic stability of ML square and triangular Fe, square $Fe_1C_1$ and $Fe_2C_2$. For structures of ML square and triangular Fe, and square $Fe_1C_1$, there are imaginary frequencies in the phonon dispersion curves (not shown). So, isolated Fe ML and square $Fe_1C_1$ ML are dynamically unstable. However, for ML $Fe_2C_2$ there is no imaginary frequency in the phonon dispersion curves and hence it is dynamically stable. The simulated phonon dispersion curves of $Fe_2C_2$ are shown in Fig. 5d.

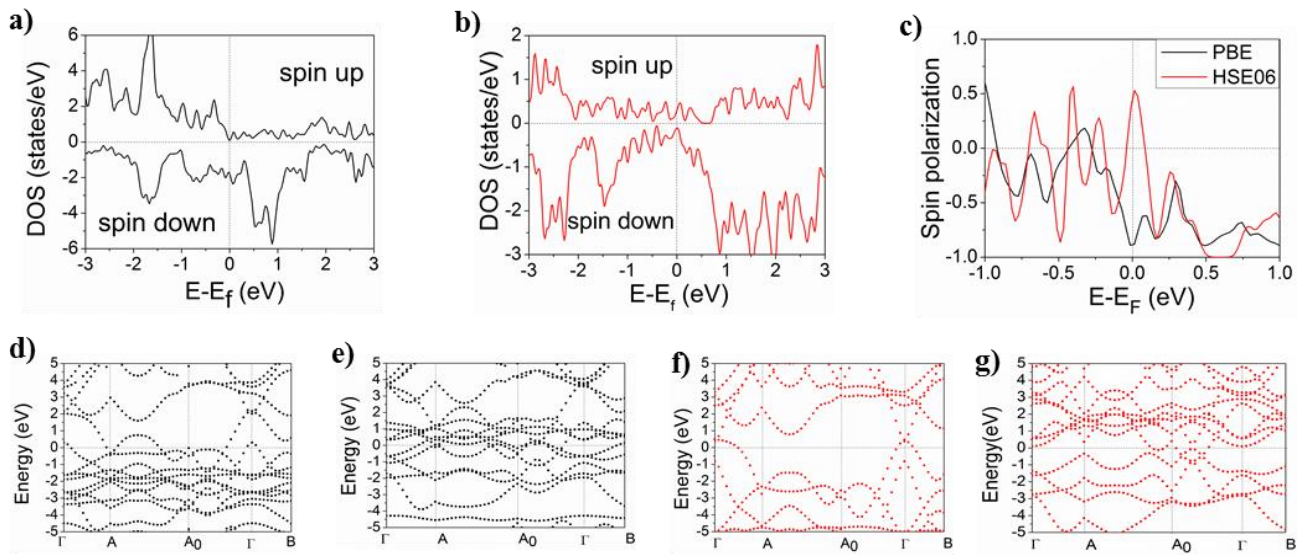

Figure 6. DOS of $Fe_2C_2$ with 0.05eV Gaussian broadening from PBE (a) and HSE06 functionals (b). Spin polarization of $Fe_2C_2$ (c). Spin-up (d) and spin-down (e) energy bands by PBE functional, and spin-up (f) and spin-down (g) band structures by HSE06 hybrid functional.

The spin polarized DOS and energy bands of $Fe_2C_2$ were calculated by both PBE and HSE06 functionals. The calculated energy bands from PBE are shown in Figure 6d and 6e. The structure of $Fe_2C_2$ remains metallic. The HSE06 energy bands are shown in Figure 6f and 6g. In Fig. 6f, only a small number of energy bands cross Fermi level. There is a strong spin polarization of 50% to 90% around Fermi level for $Fe_2C_2$ (see Fig. 6c). The spin magnetic moment of Fe is 2.15 $\mu_B$ and 2.89 $\mu_B$ per Fe atom from PBE and HSE06 calculations, respectively. We tested the magnetic moment of bulk Fe using PBE and HSE06 functionals. The spin moment of 2.26 $\mu_B$ calculated from PBE is closer to the experimental value of 2.2 $\mu_B$ than the calculated value of 2.88 $\mu_B$ from HSE06. As mentioned previously, the spin magnetic moment of pure ML Fe becomes large compared to bulk Fe. However, the introduction of carbon atoms decreases iron spin magnetic moments in Fe-C compounds. It is a combination of changed dimension (increase spin magnetic moment) and carbon interaction (decrease spin magnetic moment). Fe atoms of $Fe_2C_2$ are in medium-spin states from PBE calculations. Bader charge analyses are used to quantify the charge transfer between Fe and C; per iron atom lose 0.72 electrons.

## ■ SUMMARY

We have studied the stability, electronic and magnetic properties of Fe and Fe-C monolayers by ab initio calculations in combination with PSO algorithm. For pure Fe embedded across Gr pores, most of the Fe atoms with initial square-lattice change to triangular-lattice after structural relaxation. So Fe square-lattice is energetically unstable in graphene pores. In order to explain the experiment observed square-lattices, and to take fairly complete structures with various Fe to C ratios into consideration, we searched two dimensional Fe-C compounds with different Fe concentrations using the PSO technique. A $Fe_1C_1$ structure in square-lattice embedded in graphene is found and Fe can maintain at its square-lattice sites after relaxation. The bond lengths of Fe-Fe in the Fe-C compound fit to that in TEM experiment very well, in the range from about 2.45 to 2.85 Å; while that of pure monolayer Fe is much smaller, only about 2.3 Å. More importantly, a $Fe_2C_2$ "quasicrystal" structure is predicted and is dynamic stable. The total energy of $Fe_2C_2$ is lower than $Fe_1C_1$ by 0.83 eV per $Fe_1C_1$. The $Fe_2C_2$ is the most stable than both $Fe_1C_1$ and the separate systems of ML Fe and Gr. Fe atoms themselves in $Fe_2C_2$ are in a close-to-square lattice. Based on the above analysis of lattice size, lattice shape and total energies, we conclude that the square lattice in Gr observed in TEM images may be a mixture of (or something between) $Fe_2C_2$ and $Fe_1C_1$. The spin magnetic moment of Fe decreases due to the charge transfer between Fe and C; while large spin-polarization around the Fermi Level of different Fe-C structures are found, which could benefit for spintronic applications. Other properties of $Fe_2C_2$, such as its mechanical properties including mechanical stability, and thermal stability, are under study.

**ACKNOWLEDGMENTS:** This work is supported by National Natural Science Foundation of China (Grants 11474145 and 11334003). We thank the National Supercomputing Center in Shenzhen for providing computation time.

**TOC Graphic:**

## Fe square-lattices in graphene pores

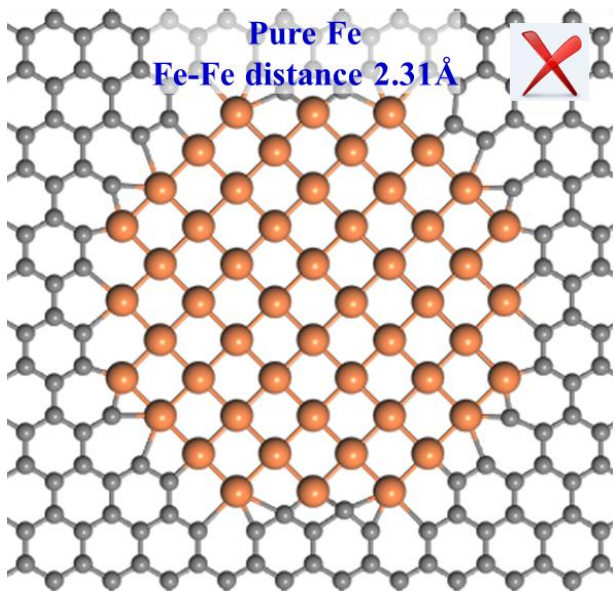 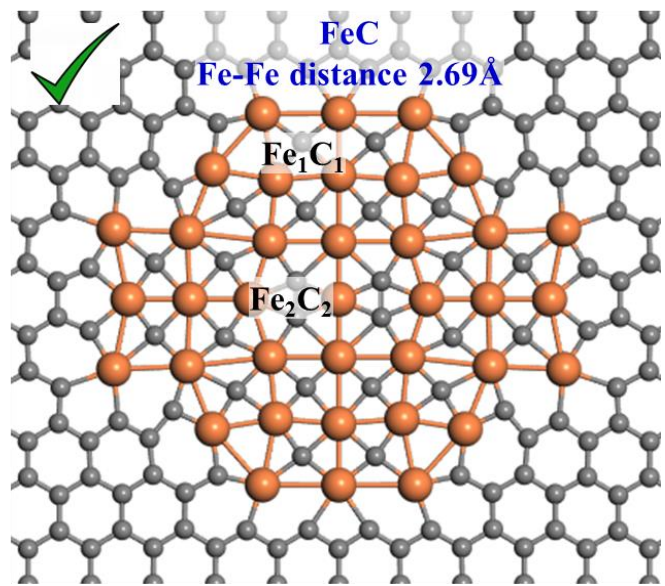